\documentclass[twoside]{article}
\usepackage{fleqn,espcrc2}

\usepackage{graphicx}
\usepackage{epsfig}
\usepackage[figuresright]{rotating}

\newcommand{\AmS}{{\protect\the\textfont2
  A\kern-.1667em\lower.5ex\hbox{M}\kern-.125emS}}

\def\VEV#1{\left\langle #1\right\rangle}

\def\ifmath#1{\relax\ifmmode #1\else $#1$\fi}
\def\ls#1{\ifmath{_{\lower1.5pt\hbox{$\scriptstyle #1$}}}}
\def\ph{\phantom{e}}
\def\anti{\overline}
\def\call{{\cal L}}
\def\epem{e^+e^-}

\def\gam{\gamma}
\def\lam{\lambda}
\def\MSUSY{M_{\rm SUSY}}
\def\MSUSYY{M_{\rm SUSY}^2}
\def\mstopa{M_{\widetilde t_1}}
\def\mstopb{M_{\widetilde t_2}}
\def\msusy{M_{\rm SUSY}}

\def\gtino{\wt g_{3/2}}
\def\mgtino{m_{3/2}}

\def\hsm{h_{\rm SM}}

\def\hl{h^0}
\def\hh{H^0}
\def\ha{A^0}
\def\hp{H^+}
\def\hm{H^-}
\def\hpm{H^{\pm}}
\def\mhl{m_{\hl}}
\def\mhh{m_{\hh}}
\def\mha{m_{\ha}}

\def\mhpm{m_{\hpm}}
\def\mhmax{m_{h^0}^{\rm max}}
\def\tanb{\tan\beta}

\def\mt{m_t}

\def\mz{m_Z}
\def\mw{m_W}
\def\mx{M_X}
\def\rts{\sqrt s}
\def\etmiss{E_T^{\rm miss}}

\def\pbi{~{\rm pb}^{-1}}

\def\gev{~{\rm GeV}}
\def\tev{~{\rm TeV}}

\def\gl{\wt g}

\def\staur{\wt \tau_R}

\def\mstaur{m_{\staur}}
\def\snu{\wt\nu}

\def\cnone{\wt\chi^0_1}
\def\cntwo{\wt\chi^0_2}

\def\cpmone{\wt \chi^{\pm}_1}

\def\mcpmone{m_{\cpmone}}

\def\mcnone{m_{\cnone}}

\def\sq{\wt q}

\def\slep{\wt \ell}

\def\sbot{\wt b}

\def\china{\widetilde\chi^0_1}
\def\chinb{\widetilde\chi^0_2}

\def\chipa{\widetilde\chi^+_1}

\def\mpl{M_{\rm P}}
\def\wt{\widetilde}
\def\wh{\widehat}
\def\lsim{\mathrel{\raise.3ex\hbox{$<$\kern-.75em\lower1ex\hbox{$\sim$}}}}
\def\gsim{\mathrel{\raise.3ex\hbox{$>$\kern-.75em\lower1ex\hbox{$\sim$}}}}

\def\eq#1{eq.~(\ref{#1})}
\def\eqs#1#2{eqs.~(\ref{#1}) and (\ref{#2})}
\def\Ref#1{ref.~\cite{#1}}

\def\fig#1{fig.~\ref{#1}}

\def\beq{\begin{equation}}
\def\eeq{\end{equation}}
\newenvironment{Eqnarray}%
     {\arraycolsep 0.14em\begin{eqnarray}}{\end{eqnarray}}
\def\beqa{\begin{Eqnarray}}
\def\eeqa{\end{Eqnarray}}
\def\nicefrac#1#2{\hbox{${#1\over #2}$}}
\def\half{\nicefrac{1}{2}}
\def\threehalf{\nicefrac{3}{2}}
\def\fivethirds{\nicefrac{5}{3}}
\def\lp{\lambda^\prime}
\def\ie{{\it i.e.}}
\def\eg{{\it e.g.}}

\def\etal{{\it et al.}}
\def\pri{^{\prime}}
\let\us=\underline
\let\ds=\displaystyle
\def\MPL #1 #2 #3 {{\sl Mod.~Phys.~Lett.}~{\bf#1} (#3) #2}
\def\NPB #1 #2 #3 {{\sl Nucl.~Phys.}~{\bf B#1} (#3) #2}
\def\PLB #1 #2 #3 {{\sl Phys.~Lett.}~{\bf B#1} (#3) #2}
\def\PR #1 #2 #3 {{\sl Phys.~Rep.}~{\bf#1} (#3) #2}
\def\PRD #1 #2 #3 {{\sl Phys.~Rev.}~{\bf D#1} (#3) #2}
\def\PRL #1 #2 #3 {{\sl Phys.~Rev.~Lett.}~{\bf#1} (#3) #2}
\def\RMP #1 #2 #3 {{\sl Rev.~Mod.~Phys.}~{\bf#1} (#3) #2}
\def\ZPC #1 #2 #3 {{\sl Z.~Phys.}~{\bf C#1} (#3) #2}
\def\IJMP #1 #2 #3 {{\sl Int.~J.~Mod.~Phys.}~{\bf#1} (#3) #2}

\title{Low-Energy Supersymmetry and its Phenomenology}

\author{Howard E. Haber\address{Santa 
        Cruz Institute for Particle Physics \\ 
        University of California, Santa Cruz, CA 95064, USA}%
        \thanks{Supported in part by the U.S. Department of Energy}}
       
\begin{document}

\begin{abstract}
The structure of low-energy supersymmetric models of
fundamental particles and interactions is reviewed, with 
an emphasis on the minimal supersymmetric extension of the 
Standard Model (MSSM) and some of its variants.  Various
approaches to the supersymmetry-breaking mechanism are considered.
The implications for the phenomenology of Higgs bosons and 
supersymmetric particles at future colliders are discussed.
\vspace{1pc}
\end{abstract}

\maketitle
\setcounter{footnote}{0}

\section{INTRODUCTION}

The Standard Model provides a remarkably successful description of the
properties of the quarks, leptons and spin-1 gauge bosons at energy scales
of ${\cal O}(100)$~GeV and below.  However, the Standard Model is not the
ultimate theory of the fundamental particles and their interactions.
At an energy scale above the Planck scale, $\mpl\simeq 10^{19}$~GeV, quantum
gravitational effects become significant
and the Standard Model must
be replaced by a more fundamental theory that incorporates
gravity.
It is also possible that the Standard Model breaks down at some 
energy scale, $\Lambda$, below the Planck scale.
In this case, the Standard Model degrees of freedom are no longer
adequate for describing the physics above $\Lambda$ and new physics
must enter. 
Thus, the Standard Model is not a {\it fundamental} theory;
at best, it is an {\it effective field theory}~\cite{EFT}.
At an energy scale below $\Lambda$, the Standard Model (with
higher-dimension operators to parameterize the new physics at the scale
$\Lambda$) provides an extremely good description of all observable
phenomena.

In an effective field theory, all parameters of the low-energy theory
({\it i.e.} masses and couplings) are calculable in terms of parameters
of a more fundamental, renormalizable theory that
describes physics at the energy scale $\Lambda$.  All
low-energy couplings and fermion masses are logarithmically sensitive to
$\Lambda$.  In contrast, scalar squared-masses are {\it quadratically}
sensitive to $\Lambda$.  The Higgs
mass (at one-loop) has the following form:
\beq \label{natural}
m_h^2= (m_h^2)_0+{cg^2\over 16\pi^2}\Lambda^2\,,
\eeq
where $(m_h^2)_0$ is a parameter of the fundamental theory and $c$ is a
constant, presumably of ${\cal O}(1)$, which is calculable within the
low-energy effective theory.  The ``natural'' value
for the scalar squared-mass is $g^2\Lambda^2/16\pi^2$.  Thus, the expectation
for $\Lambda$ is
\beq \label{tevscale}
\Lambda\simeq {4\pi m_h\over g}\sim {\cal O}(1~{\rm TeV})\,.
\eeq
If $\Lambda$ is significantly larger than 1~TeV (often called the
hierarchy and naturalness problem in the literature~\cite{naturally}),
then the only way
for the Higgs mass to be of order the scale of electroweak symmetry
breaking is to have an ``unnatural'' cancellation between the two terms
of \eq{natural}.
This seems highly unlikely given that the two terms of
\eq{natural} have completely different origins.  

A viable theoretical framework that incorporates weakly-coupled Higgs
bosons and satisfies the constraint of \eq{tevscale} is that of
``low-energy'' or ``weak-scale'' supersymmetry \cite{Nilles84,Haber85,smartin}.
In this framework, supersymmetry is
used to relate fermion and boson masses and interaction strengths.  
Since fermion masses are only
logarithmically sensitive to $\Lambda$, boson masses will exhibit the
same logarithmic sensitivity if supersymmetry is exact.  Since no
supersymmetric partners of Standard Model particles have yet been
found, supersymmetry cannot be an exact symmetry of nature.
Thus, $\Lambda$ should be identified 
with the supersymmetry-breaking scale.  The
naturalness constraint of \eq{tevscale} is still relevant, so in the
framework of low-energy supersymmetry, the scale of supersymmetry
breaking should not be much larger than about 1~TeV in order that the
naturalness of scalar masses be preserved.  The supersymmetric extension
of the Standard Model would then replace the Standard Model as the
effective field theory of the TeV scale.  
One advantage of the supersymmetric
approach is that the effective low-energy supersymmetric theory {\it
can} be valid all the way up to the Planck scale, while still being
natural!  The unification of the three gauge couplings at an energy
scale close to the Planck scale, which does not occur in the Standard
Model, is seen to occur in the minimal supersymmetric extension of the
Standard Model, and provides an additional motivation for seriously 
considering the low-energy supersymmetric framework~\cite{susyguts}.

\section{STRUCTURE OF THE MSSM}

The minimal supersymmetric extension of the Standard Model (MSSM)
consists of taking the Standard Model
and adding the corresponding supersymmetric partners.
(For a review of the structure of the MSSM, see \eg,
refs.~\cite{Haber85,smartin}.)
In addition, the MSSM contains two hypercharge $Y=\pm 1$ Higgs
doublets ($H_u$ and $H_d$, respectively),
which is the minimal structure for the Higgs sector of an
anomaly-free supersymmetric extension of the Standard Model.
The supersymmetric structure of the theory also requires (at least) two
Higgs doublets to generate mass for both ``up''-type and ``down''-type
quarks (and charged leptons) \cite{Inoue82,Gunion86}.
The supersymmetric spectrum can be described by a set of superfields
that incorporates both the Standard Model fields and their superpartners
as shown in Table~\ref{superfields}.
\renewcommand{\arraystretch}{1.3}
\setlength{\tabcolsep}{0.01in}
\begin{table}[htb]
\centering
\caption{The MSSM Particle Spectrum}
\vskip6pt
\begin{tabular}{ccc}
          &             & Fermionic \\[-5pt]
Superfield& Boson Fields& Partners \\
\hline
\multicolumn{2}{l}{\us{Gauge Multiplets}}& \\[3pt]
$\wh G$&  $g$&     $\wt g$ \\
$\wh V^a$&$W^a$&  $\widetilde W^a$ \\
$\wh V\pri$&     $B$&      $\widetilde B$ \\
\hline
\multicolumn{2}{l}{\us{Matter Multiplets}} & \\[3pt]
$\ds{\wh L\atop \wh E}$&
 leptons $\Bigg\{ \ds{\wt L^j\,=\,(\widetilde\nu,\widetilde e^-)_L  \atop
           \ds{\wt E\,=\,\widetilde e^+_R\hphantom{(\nu,_L)}}}$\hfill&
            $\ds{ (\nu,e^-)_L \atop  e^c_L}$
\\[16pt]
$\ds{\wh Q\atop \ds{\wh U \atop \wh D} } $&
  quarks $\left\{\vbox to 27pt{}   \right.
 \ds{ \wt Q^j\,=\,(\widetilde u_L,\widetilde d_L)
  \atop \ds{\wt U\,=\,\widetilde u^*_R\hphantom{,d_L)^f}
  \atop \ds{\wt D\,=\,\widetilde d^*_R\hphantom{,d_L)^f}} } }$\hfill&
    $\ds{(u,d)_L \atop\ds{u^c_L \atop d^c_L}}$
\\ \noalign{\vskip8pt}
$\ds{\wh H_d\atop \wh H_u}$&
     Higgs $\Bigg\{ \ds{ H^i_d \atop \ds{H^i_u}}$\hbox to 2.0cm{}&
    $\ds{(\wt H^0_d,\wt H^-_d)_L \atop (\wt H^+_u,\wt H^0_u)_L}$
\\ \noalign{\vskip8pt}
\hline
\end{tabular}
\label{superfields}
\end{table}
 Note that
Table~\ref{superfields} 
lists the interaction eigenstates. Particles with the same
SU(3)$\times$U(1)$_{\rm EM}$ quantum numbers can mix.  The physical
(mass) eigenstates are determined from the full interaction Lagrangian
of the theory.  For example, the physical {\it charginos}
[$\widetilde\chi^\pm_j$ ($j=1,2$)] are linear combinations of the
charged winos ($\wt W^\pm$) and charged higgsinos
($\wt H_u^+$, $\wt H_d^-$), while the physical
{\it neutralinos} [$\widetilde\chi^0_k$
($k=1,\ldots,4$)] are linear combinations of the
neutral wino ($\wt W^3$), bino ($\wt B$) and neutral higgsinos 
($\wt H_u^0$, $\wt H_d^0$).

The renormalizable supersymmetric interactions
are fixed once the superpotential (a cubic polynomial of superfields)
is chosen.  In the MSSM, the most general SU(3)$\times$SU(2)$\times$U(1)
invariant cubic superpotential is given by $W\equiv W_{\rm RPC}+
W_{\rm RPV}$, where
\beqa \label{wrparity}
  W_{\rm RPC}
& = &(y_e)_{mn} \widehat H_d \widehat L_m
      \widehat E_n + (y_d)_{mn}\widehat H_d \widehat Q_m\widehat D_n
       \nonumber \\
     &&\qquad -  (y_u)_{mn} \widehat H_u \widehat Q_m\widehat U_n
       - \mu \widehat H_d \widehat H_u\,,
\eeqa
and
\beqa \label{nrsuppot}
W_{\rm RPV} & = & 
(\lambda_L)_{mnp}\widehat L_m \widehat L_n
\widehat E_p + (\lp_L)_{mnp} \widehat L_m \widehat Q_n
\widehat D_p \nonumber \\
&&  +(\lambda_B)_{mnp}\widehat U_m \widehat D_n \widehat D_p
-(\mu\ls{L})_n \widehat L_n \widehat H_u \,.
\eeqa
In \eqs{wrparity}{nrsuppot}, $m$, $n$ and $p$ label the generation
number, and SU(2) doublet fields are contracted by the invariant
antisymmetric tensor $\epsilon_{ij}$ (with $\epsilon_{12}=1$).
For example, the Yukawa interactions are obtained
from the trilinear terms of \eq{wrparity} by
taking all possible combinations involving two fermions
and one scalar superpartner.
The gauge multiplets couple to matter multiplets in a manner consistent
with supersymmetry and the SU(3)$\times$SU(2)$\times$U(1) gauge
symmetry.

In contrast to the Standard Model, the terms of dimension $\leq 4$
in the supersymmetric Lagrangian
do not automatically preserve baryon number (B) and lepton
number (L).  In order to be consistent with the
(approximately) conserved baryon number and lepton number observed in
nature, one must impose constraints on the terms in the superpotential
[\eqs{wrparity}{nrsuppot}].  In particular, the terms of $W_{\rm RPC}$
conserve B and L, whereas the terms of $W_{\rm RPV}$ 
violate either B or L as indicated by the subscript of the corresponding
coefficients.  If all terms in $W_{\rm RPV}$ were present, some of the
coefficients would have to be extremely small in order to
avoid proton decay at a catastrophic rate.

In the MSSM, one enforces an approximate lepton and baryon number
invariance by imposing R-parity-invariance.
Any particle of spin $S$, baryon number $B$ and lepton number $L$
then possess a multiplicatively-conserved
R-parity given by $R=(-1)^{3(B-L)+2S}$.  
Then, all operators in the R-parity-conserving (RPC) model
of dimension $\leq 4$ exactly conserve B and L.
In the RPC model, R-parity conservation is equivalent to imposing a ${\bf Z_2}$
matter parity on the superpotential according to the quantum numbers
specified in Table~\ref{parity}, which results in $W_{\rm RPV}=0$.
Thus, the supersymmetric interactions of the MSSM
depend on the following parameters:
\begin{itemize}
\item
 SU(3)$\times$SU(2)$\times$U(1) gauge couplings, $g_3$, $g_2\equiv g$,
and $g_1\equiv\sqrt{\fivethirds}\,g^\prime$;
\item
(complex) Higgs-fermion Yukawa coupling matrices, ${\bf y_u}$,
${\bf y_d}$, and ${\bf y_e}$; and
\item
a supersymmetric Higgs mass parameter, $\mu$.
\end{itemize}
Since the MSSM is a model of three generations of quarks, leptons and
their superpartners, ${\bf y_u}$, ${\bf y_d}$, and ${\bf y_e}$ are
complex $3\times 3$ matrices.  
However, not all these degrees of
freedom are physical, as emphasized below.

Note that in the (minimal version of the) Standard Model and in the MSSM,
neutrinos are exactly massless if only terms of dimension $\leq 4$ in
the Lagrangian are considered.  Majorana neutrino masses may be
generated in either theory by introducing the appropriate dimension-5
gauge invariant $\Delta L=2$ terms.  In the MSSM,
such terms can be realized
via an RPC supersymmetric extension of the seesaw mechanism,
which would require the addition of a 
SU(3)$\times$SU(2)$\times$U(1) singlet superfield~\cite{susyseesaw}.

\renewcommand{\arraystretch}{1.3}
\setlength{\tabcolsep}{0.01cm}
\begin{table}[t]
\centering
\caption{Matter discrete symmetries}
\vskip3pt
\begin{tabular}{cccccccc}
symmetry\ph &\ph $\widehat Q_n$\ph & \ph $\widehat U_n$ \ph& \ph 
$\widehat D_n$ \ph & \ph $\widehat L_n$\ph &\ph $\widehat E_n$\ph &\ph 
$\widehat H_U$\ph &\ph $\widehat H_D$\ph \\
\hline
${\bf Z_2}$ & $-1$ & $-1$ & $-1$& $-1$ & $-1$ & $+1$ & $+1$ \\
${\bf Z_3}$ &  $\omega$ & $\omega^{-1}$  &  $\omega^{-1}$ & 
$+1$ & $+1$ & $+1$ & $+1$ \\
\hline
$\omega\equiv e^{\,2i\pi/3}$ \\
\label{parity}\end{tabular}
\vspace{-0.5in}
\end{table}

One can also avoid B violation in an R-parity-violating (RPV) theory
by imposing a discrete ${\bf Z_3}$ triality shown in Table~\ref{parity}.
This is the unique
choice for a (generation-independent) discrete symmetry 
with no discrete gauge anomalies in a model consisting only of the
MSSM superfields~\cite{ibanez}.  
In particular, the ${\bf Z_3}$ discrete symmetry requires that
$\lambda_B=0$ in \eq{nrsuppot} [in fact, all operators of dimension 
$\leq 5$ preserve B], while all other terms in the
superpotential, including all possible L-violating terms,
are permitted.  Hence, one must check that the magnitude of the
L-violating interactions is consistent with experimental
bounds on L-violating processes~\cite{dedes}.  One good feature of 
this model is that it provides a mechanism for non-zero neutrino
masses without requiring the introduction of a new superfield beyond
those already contained in the MSSM.

At this stage, one does not yet have a realistic model of fundamental
particles and their interactions, since supersymmetry (SUSY) is unbroken.
However, the fundamental origin of SUSY-breaking
is not known at present.  Without a fundamental theory of
SUSY-breaking, the best we can do is to parameterize our 
ignorance and introduce the
most general renormalizable soft SUSY-breaking
terms \cite{Girardello82}
consistent with the SU(3)$\times$SU(2)$\times$U(1) gauge symmetry and
any additional discrete matter symmetries
that have been imposed.  
It is here where most of the new supersymmetric
model parameters reside.\footnote{It was argued in \Ref{gensusy}
that additional dimension-3 SUSY-breaking terms, which were {\it not}
characterized as ``soft'' by \Ref{Girardello82}, should be considered
in models such as the MSSM that contain no gauge-singlet superfields.}
If super\-symmetry is relevant
for explaining the scale of electroweak interactions, then the
mass parameters introduced by the soft SUSY-breaking
terms must be of order 1~TeV or below \cite{Barbieri88}.

In the MSSM, the soft-SUSY-breaking terms
consist of~\footnote{Here, we omit the more general dimension-3
SUSY-breaking terms advocated in \Ref{gensusy}.  In addition, we note
that dimension-4 SUSY-breaking terms can also be generated by
high-energy-scale physics, 
albeit with very small coefficients~\cite{Martin:2000hc}.}
(following the notation of \Ref{smartin}):
\beqa \label{softsusy}
 &&\!\!\! V_{\rm soft}  =  m^2_{H_d} |H_d|^2 \!+\! m^2_{H_u}|H_u|^2
\!-\!
                 (b H_d H_u + {\rm h.c.}) \nonumber
\\
 &&+ (m^2_{Q})_{mn}\,\widetilde Q^{i*}_m
        \widetilde Q^i_n
   +  (m^2_{\overline u})_{mn}\,\widetilde U_m^*\widetilde U_n
          \nonumber \\
 &&+  (m^2_{\overline d})_{mn}\,\widetilde D_m^*\widetilde D_n
            \nonumber \\
 && +  (m^2_{L})_{mn}\,\widetilde L^{i*}_m\widetilde L^i_n
      + (m^2_{\overline e})_{mn}\,\widetilde E_m^*\widetilde E_n
            \nonumber \\
 && +  \bigl[ (a_e)_{mn} \widetilde H_d
       \widetilde L_m \widetilde E_n\! +\! (a_d)_{mn}
       \widetilde H_d\widetilde Q_m\widetilde D_n \nonumber \\
 &&\quad\qquad - (a_u)_{mn} \widetilde H_u
        \widetilde Q_m\widetilde U_n + {\rm h.c.}\bigr]\nonumber \\
 && +  \half \left[ M_3\, \widetilde g
   \,\widetilde g + M_2 \widetilde W^a\widetilde W^a
  + M_1 \widetilde B \widetilde B +{\rm h.c.}\right]\,.
\eeqa
In the RPV model,
additional $A$ and $b$ parameters can be added, corresponding to the
additional terms that appear in the superpotential.\footnote{In the
literature, a different matrix $A$-parameter is often defined via
the relation: ${\bf a_f}\equiv {\bf y_f A_f}$
[${\bf f}={\bf u}$, ${\bf d}$, ${\bf e}$], where the ${\bf y_f}$ are the
Higgs-fermion Yukawa coupling matrices, and a $B$-parameter is defined
by $b\equiv \mu B$, where $\mu$ is the supersymmetric Higgs mass parameter.} 

The Higgs scalar potential receives contributions from both the
SUSY-conserving and the SUSY-breaking sector.
The SU(2)$\times$U(1) electroweak symmetry is spontaneously broken only
when the terms of $V_{\rm soft}$ [\eq{softsusy}] are included.  The
neutral Higgs fields acquire vacuum expectation values:
$\VEV{H_d^0}=v_d/\sqrt{2}$ and $\VEV{H_u^0}=v_u/\sqrt{2}$, where
$v^2\equiv v_d^2+v_u^2=(246~\gev)^2$ is determined by the $Z$ mass,
and $\tanb\equiv v_u/v_d$ is a free parameter.   

Not all of the parameters appearing in \eqs{wrparity}{softsusy}
represent independent degrees of freedom.  By suitable
redefinitions of the various fields of the model, one can remove all
unphysical degrees of freedom and identify the correct number of
physical parameters of the model~\cite{sutter}.  For example, the  
MSSM is characterized by  124 independent parameters, of which 18
correspond to Standard Model parameters (including $\theta_{\rm QCD}$),
one corresponds to a Higgs
sector parameter (the analogue of the Standard Model
Higgs mass), and 105 are genuinely new parameters of the model.
All together, among the various complex parameters of the model, there
are 45 phases that cannot be removed; the remaining 79 real parameters
consist of masses, real couplings and mixing angles.
Thus, an appropriate name for the minimal supersymmetric extension of
the Standard Model described above is MSSM-124 \cite{habersusy97}.
In Table~\ref{count}, we compare the
parameter count of the Standard Model with three possible versions of
low-energy supersymmetry based on the fields of the 
MSSM.  The RPC model is denoted in Table~\ref{count} simply by
MSSM, whereas the RPV model with a ${\bf Z_3}$ triality that
preserves B is denoted by (MSSM)$_{\rm B}$.  Finally, (MSSM)$_{\rm RPV}$ 
denotes the most general RPV model in which all terms in $W_{\rm RPV}$
[\eq{nrsuppot}] and the corresponding soft SUSY-breaking
terms are allowed.
\renewcommand{\arraystretch}{1.3}
\setlength{\tabcolsep}{0.1in}
\begin{table}[t]
\centering
\caption{Parameter Count in the Standard Model and MSSM Extensions}
\vskip6pt
\begin{tabular}{c|c|c|c}
model &\ real\phantom{e} & phases & \ TOTAL \\
\hline
SM\phantom{e} & 17& 2 & 19 \\
MSSM & 79 & 45 &124 \\
(MSSM)$_{\rm B}$ & 157 & 122 &279 \\
(MSSM)$_{\rm RPV}$ & 175 & 140 & 315 \\
\hline
\end{tabular}
\label{count}
\end{table}

Even in the absence of a fundamental theory of 
SUSY-breaking, one is hard-pressed to regard MSSM-124 as a fundamental
theory.  For example, no fundamental
explanation is provided for the origin of electroweak symmetry breaking.
Moreover, MSSM-124 is
not a phenomenologically viable theory over most of its parameter space.
Among the phenomenological deficiencies
are: (i) no conservation of the separate lep\-ton numbers
L$_e$, L$_\mu$, and L$_\tau$; (ii) unsuppressed
flavor-changing neutral currents (FCNC's) \cite{masiero}; and (iii)
new sources of CP-violation that are
inconsistent with the experimental bounds \cite{edmlimits,masiero2}.  As a
result, almost the entire MSSM-124 parameter space is ruled out!
This theory is viable only at very special ``exceptional'' points of
the full parameter space.

A truly Minimal SSM does not (yet) exist.  In the present usage, the
word ``minimal'' in MSSM refers to the minimal particle spectrum and
the associated R-parity invariance.  The MSSM
particle content must be supplemented by assumptions about the origin of
SUSY-breaking that lie outside the low-energy domain of the
model.  Moreover, a comprehensive map of the
phenomenologically acceptable region of MSSM-124 parameter space does
not yet exist. This
presents a formidable challenge to supersymmetric particle searches that
must impose some parameter constraints while trying to ensure that the
search is as complete as possible.  Ultimately, the goal of any
supersymmetric particle search is to measure as many of the 124
parameters as is feasible and to determine any additional
parameters that would characterize a possible departure from the minimal
SSM structure.

\section{REDUCING THE MSSM PARAMETER FREEDOM}

There are two general approaches for reducing the parameter freedom of
MSSM-124. In the low-energy (or ``bottom-up'') approach,
an attempt is made to elucidate the nature of
the exceptional points in the MSSM-124 parameter space that are
phenomenologically viable.  Consider the following two possible choices.
First, one can assume that the squark and slepton squared-mass matrices
and the matrix $A$-parameters, ${\bf A_f}$,
are proportional to the $3\times 3$ unit matrix
(horizontal universality \cite{4b,georgi,sutter}).
Alternatively, one can simply
require that all the aforementioned matrices are flavor diagonal in a
basis where the quark and lepton mass matrices are diagonal (flavor
alignment \cite{seibergnir}).  In these approaches,
the number of free parameters characterizing the MSSM
is substantially less than 124.  Moreover, L$_e$, L$_\mu$ and
L$_\tau$ are separately
conserved, while tree-level FCNC's are automatically absent.
The resulting models 
are phenomenologically viable, although there is no strong
theoretical basis for either approach.

In the high-energy (or ``top-down'') approach, 
the MSSM-124 parameter freedom is reduced by imposing theoretical
constraints on the structure of SUSY-breaking.
It is very difficult (perhaps impossible) to construct a model of
low-energy super\-symmetry where the
SUSY-breaking arises solely as a consequence of
the interactions of the particles of the
MSSM.  A more viable scheme posits a theory
in which the fundamental source of SUSY-breaking originates in
a sector that is distinct from the fields that make up the MSSM.
For lack of a better term, we will call this new sector
the direct SUSY-breaking (DSB) sector.\footnote{In the
literature, the DSB-sector
usually refers to the sector of dynamical SUSY-breaking.
Here, we shall interpret the word {\it dynamical} in its broadest
sense.  Dynamical SUSY-breaking can be non-perturbative in nature
(\eg, gaugino condensation \cite{gcondense}) or perturbative in nature.
Examples of the latter include tree-level O'Raifeartaigh (or $F$-type)
breaking and Fayet-Iliopoulos (or $D$-type) breaking.  For additional
details, see \Ref{smartin}.}
The SUSY-breaking
inherent in the DSB-sector is subsequently transmitted to the
MSSM spectrum by some mechanism.

Integrating out the physics associated with the DSB sector, one
obtains initial conditions for the MSSM parameters at some high energy
scale.  Renormalization group (RG) evolution then allows one to 
evolve down to the electroweak scale and derive the
full MSSM particle spectrum.  One bonus of this analysis is 
that one of the diagonal Higgs squared-mass
parameters is typically driven negative by the
RG-evolution.  Thus, electroweak symmetry breaking is generated
radiatively, and the resulting electroweak symmetry-breaking scale is
intimately tied to the scale of low-energy SUSY-breaking.
 
Two theoretical scenarios have been examined in detail: gravity-mediated
and gauge-mediated SUSY-breaking. In both these approaches,
supersymmetry is spontaneously broken, in which case a massless Goldstone
fermion, the {\it goldstino}, arises.  Its coupling to a particle and
its superpartner is fixed by the supersymmetric Goldberger-Treiman
relation \cite{sgtrel} 
\beq
\call_{\rm int}=-{1\over F}\,j^{\mu\alpha}\partial_\mu \wt G_\alpha+{\rm
h.c.}\,,
\label{gtrel}
\eeq
where $j^{\mu\alpha}$ is the supercurrent, which depends bilinearly on
all the fermion--boson superpartner pairs of the theory and
$\wt G_\alpha$ is the spin-1/2 goldstino field (with spinor index
$\alpha$).
In particular, $\sqrt{F}$ is the scale of
direct SUSY-breaking
which occurs in the DSB-sector (typically, $\sqrt{F}\gg\mz$).
When gravitational effects are included, the goldstino is ``absorbed''
by the {\it gravitino} ($\gtino $),
the spin-3/2 partner of the graviton.
By this super-Higgs mechanism \cite{Deser77}, 
the goldstino is removed from the
physical spectrum and the gravitino acquires a mass ($m_{3/2}$).
In models where the gravitino mass is generated at tree-level (see, \eg,
\Ref{Lahanas} for further discussion), one finds:
\begin{equation}
m_{3/2}={F\over \sqrt 3\mpl}\,,
\label{mgravitino}
\end{equation}
where $\mpl$ is the reduced Planck mass.  The helicity $\pm\half$
components
of the gravitino behave approximately like the goldstino, whose
couplings to particles and their superpartners are determined by
\eq{gtrel}.  In particular, the goldstino couplings are enhanced by
a factor of $\mpl^2/F$ relative to couplings of gravitational
strength.  In contrast, the helicity $\pm\threehalf$ components
of the gravitino always couple with gravitational strength to particles
and their superpartners, and thus can be neglected in phenomenological
studies.

In many models, the DSB-sector is comprised of fields that
are completely neutral with respect to the Standard Model gauge group.
In such cases, the DSB-sector is also called the ``hidden sector.''
The fields of the MSSM are said to reside in
the ``visible sector,'' and the model is constructed such
that no renormalizable tree-level interactions exist between
fields of the visible and hidden sectors.  A third sector, the
so-called ``messenger sector,'' is often employed in models to transmit
the SUSY-breaking from the hidden sector to the visible sector.
However, it is also possible to construct models in which the DSB-sector
is not strictly hidden and contains fields that are charged with respect
to the Standard Model gauge group.

\subsection{Gravity-mediated SUSY-breaking}

All particles feel the gravitational force.  In particular,
particles of the
hidden sector and the visible sector can interact via the exchange of
gravitons.  Thus, supergravity (SUGRA) models provide a natural
mechanism
for transmitting the SUSY-breaking of the hidden sector to the
particle spectrum of the MSSM. In models of gravity-mediated
SUSY-breaking, Planck-scale physics is the messenger of
SUSY-breaking \cite{Lykken,Weldon}.

In the {\it minimal} supergravity (mSUGRA) framework \cite{Nilles84},
the soft SUSY-breaking
parameters at the Planck scale take a particularly
simple form:
\beqa \label{plancksqmasses}
 &&{\bf m^2_{Q}} (\mpl) = {\bf m^2_{\anti u}}(\mpl) =
{\bf m^2_{\anti d}}(\mpl) \nonumber \\
&&\qquad\qquad\,\,\,\!  = {\bf m^2_{L}}(\mpl) = {\bf m^2_{\anti e}}(\mpl)
=  m_0^2 {\bf 1}
\,,\nonumber\\
 &&m^2_{H_u}(\mpl) = m^2_{H_d}(\mpl) = m_0^2 \,,\nonumber \\
 && {\bf a_f}(\mpl) =A_0 {\bf y_f}(\mpl)\,,\qquad {\bf f}={\bf u,d,e}\,,
\eeqa
where ${\bf 1}$ is the $3\times 3$ identity matrix in generation space.
Note that the last condition is equivalent to ${\bf A_f}(\mpl)=
A_0{\bf 1}$.  Thus, the $3\times 3$ matrices above respect horizontal
universality at the Planck scale.

In addition, the gauge couplings and gaugino mass
parameters are assumed to unify at some high energy scale, $\mx$.
(The latter condition is automatic in models of supersymmetric
grand unification, where $\mx$ is the unification scale.)
The unification relation
\beq  \label{gunif}
M_1(M_X) = M_2(M_X) = M_3(M_X) = m_{1/2}
\eeq
implies that the
low-energy gaugino mass parameters (approximately) satisfy:
\beqa \label{gauginomassrelation}
 M_3 &=& {g^2_3\over g_2^2} M_2\simeq 3.5M_2\,,\nonumber \\[5pt]
 M_1 &=& \fivethirds\tan^2\theta_W M_2\simeq 0.5M_2\,.
\eeqa

One can count the number of independent parameters in the
mSUGRA framework.  In addition to 18 Standard Model parameters
(excluding the Higgs mass), one must specify
$m_0$, $m_{1/2}$, $A_0$, and Planck-scale values for $\mu$ and
$B$-parameters (denoted by $\mu_0$ and $B_0$).  In principle,
$A_0$, $B_0$ and $\mu_0$ can be complex, although in
the mSUGRA approach, these parameters are taken (arbitrarily)
to be real.  
Renormalization group evolution is used to compute the low-energy
values of the mSUGRA parameters, which then fixes all the
parameters of the low-energy MSSM.  In particular, the two Higgs
vacuum expectation values, $v_u$ and $v_d$ 
(or equivalently, $m_Z$ and $\tan\beta$)
can be expressed as a function of the Planck-scale supergravity
parameters.  The simplest procedure
is to remove $\mu_0$ and $B_0$ in favor of $m_Z$ and $\tan\beta$ (the
sign of $\mu_0$ is not fixed in this
process).  In this case, the MSSM spectrum
and its interaction strengths are determined by five parameters:
$m_0$, $A_0$, $m_{1/2}$, $\tan\beta$, and the
sign of $\mu_0$, in addition to the 18 parameters of the Standard Model.
The requirement of radiative electroweak symmetry-breaking
imposes an additional constraint on the possible range of the mSUGRA
parameters.  In particular, one finds that $1\lsim\tanb\lsim m_t/m_b$.
In principle, one should also include the mass of the gravitino,
$m_{3/2}$ (or equivalently, $\sqrt{F}$), in the list of independent
parameters. In mSUGRA, one arranges the scale of the hidden sector
SUSY-breaking such that $\sqrt{F}\sim 3\times 10^{10}\gev$.
In this case, the gravitino mass is of
order the electroweak symmetry-breaking scale [see \eq{mgravitino}], 
while its couplings to the MSSM fields are extremely weak.
Such a gravitino would play no role in supersymmetric phenomenology at
colliders.

Recently, attention has been given to a class of supergravity models in
which \eq{gauginomassrelation} does not hold.  In models where no 
tree-level gaugino masses are generated, 
one finds a model-independent contribution to the
gaugino mass whose origin can be traced to the super-conformal
(super-Weyl) anomaly which is common to all supergravity 
models~\cite{anomalymed}.
This approach has been called {\it anomaly-mediated} 
SUSY-breaking (AMSB).
The gaugino mass parameters (in the one-loop approximation) 
are given by:
\beq
M_i\simeq {b_i g_i^2\over 16\pi^2}m_{3/2}\,,
\label{eqanom}
\eeq
where the $b_i$ are the coefficients of the MSSM gauge beta-functions
corresponding to the corresponding U(1), SU(2) and SU(3) gauge groups:
$(b_1,b_2,b_3)= (\nicefrac{33}{5},1,-3)$.
Anomaly-mediated 
SUSY-breaking also generates (approximate) flavor-diagonal squark and
slepton mass matrices.  However, in the MSSM this cannot be the sole
source of SUSY-breaking in the slepton sector,
since the latter yields negative squared-mass contributions for the
sleptons.  A possible RPV solution, involving only the superfields of the MSSM,
has been advocated in \Ref{allanach}

\subsection{Gauge-mediated SUSY-breaking}

In {\it gauge-mediated} SUSY-breaking (GMSB),
SUSY-breaking is transmitted to the MSSM via gauge forces.
A typical structure of such models involves a hidden sector
where supersymmetry is broken, a ``messenger sector'' consisting of
particles (messengers) with SU(3)$\times$SU(2)$\times$U(1) quantum
numbers, and the visible sector consisting of the fields of the
MSSM~\cite{dinenelson,gmsbreview}.
The direct coupling of the messengers to the hidden sector generates a
SUSY-breaking spectrum in the messenger sector.
Finally, SUSY-breaking 
is transmitted to the MSSM via the virtual exchange of the
messengers.  If this approach is extended to
incorporate gravitational phenomena, then supergravity effects will
also contribute to SUSY-breaking.  However, in
models of gauge-mediated SUSY-breaking, one usually chooses the
model parameters in such a way
that the virtual exchange of the messengers dominates the effects of the
direct gravitational interactions between the hidden and visible
sectors.  In this scenario,
the gravitino mass is typically in the eV to keV range, and is therefore
the LSP.  The helicity $\pm\half$ components of $\widetilde g_{3/2}$
behave approximately like the goldstino; its coupling to the particles
of the MSSM is significantly stronger than a coupling of
gravitational strength.

In the minimal GMSB approach,
there is one effective mass scale, $\Lambda$, that determines all
low-energy scalar and gaugino mass parameters through loop-effects
(while the resulting $A$-parameters are suppressed).
In order that the resulting superpartner masses
be of order 1~TeV or less, one must have $\Lambda\sim 100$~TeV.
The origin of the $\mu$ and $B$-parameters is model dependent and
lies somewhat outside the GMSB ansatz.  The simplest models of
this type are even more restrictive than mSUGRA, with two fewer
degrees of freedom.   However, minimal GMSB is not a fully realized
model.  The sector of SUSY-breaking dynamics can be very
complex, and no complete GMSB model 
yet exists that is both simple and compelling.

\section{THE MSSM HIGGS SECTOR}

\subsection{The Higgs Sector in Low-Energy Supersymmetry}

In the MSSM, the Higgs sector is a two-Higgs-doublet model
with Higgs self-interactions constrained by supersymmetry
\cite{Inoue82,Gunion86,hhg,Haber97a}. Moreover, in spite
of the large number of potential CP-violating phases among the
MSSM-124 parameters, the tree-level MSSM Higgs sector is
automatically CP-conserving.  In particular, unphysical phases can be
absorbed into the definition of the Higgs fields such that
$\tan\beta$ is real and positive.
As a result, the physical neutral Higgs scalars are CP-eigenstates.
There are five physical Higgs particles in this model: a charged Higgs
pair ($\hpm$), two CP-even neutral Higgs bosons (denoted by $\hl$
and $\hh$ where $\mhl \leq \mhh$) and one CP-odd neutral
Higgs boson ($\ha$).   At tree level, $\tan\beta$
and one Higgs mass (usually chosen to be $\mha$)
determine the tree-level Higgs-sector parameters.
These include the other Higgs masses,
an angle $\alpha$ [which measures the component of the original
$Y=\pm 1$ Higgs doublet states in the physical CP-even
neutral scalars], the Higgs boson self-couplings,
and the Higgs boson couplings to particles of the Standard Model and
their superpartners.

When one-loop radiative corrections are incorporated, the Higgs
masses and couplings depend on
additional parameters of the supersymmetric model
that enter via virtual loops.  
One of the most striking effects of the radiative corrections to the
MSSM Higgs sector is the modification of the upper bound of the
light CP-even Higgs mass, as first noted in \Ref{hhprl}.  
When $\tanb\gg 1$ and $\mha\gg\mz$, the
{\it tree-level} prediction for $\mhl$ corresponds to its theoretical
upper bound, $\mhmax=\mz$.  Including radiative corrections, the theoretical
upper bound is increased, primarily because of an incomplete
cancellation of the top-quark and top-squark (stop) loops (these
effects actually cancel in the exact supersymmetric limit).
The relevant parameters that govern the stop sector are
the average of the two stop squared-masses: $\MSUSYY\equiv
\half(\mstopa^2+\mstopb^2)$, and the off-diagonal element of the
stop squared-mass matrix: $m_t X_t\equiv m_t(A_t-\mu\cot\beta)$.
The qualitative behavior of the radiative corrections can be most easily
seen in the large top squark mass limit, where in addition the
splitting of the two diagonal entries and the off-diagonal entry
of the stop squared-mass matrix are both small in comparison to
$\MSUSYY$.  In this case, the upper bound on the lightest CP-even Higgs
mass is approximately given by
\beqa \label{deltamh}
\mhl^2 & \lsim & \mz^2+{3g^2\mt^4\over
8\pi^2\mw^2}\left[\ln\left({\MSUSYY\over\mt^2}\right)\right. \nonumber \\[5pt]
&&\left.\qquad\quad +{X_t^2\over \MSUSYY}
\left(1-{X_t^2\over 12\MSUSYY}\right)\right]\,,
\eeqa
More complete treatments of the radiative corrections
include the effects of
stop mixing, renormalization group improvement, and the
leading two-loop contributions, and imply that
\eq{deltamh} somewhat overestimates the true upper bound of 
$\mhl$ (see \Ref{higgsrad} for the most recent results).
Nevertheless, \eq{deltamh} correctly reflects some noteworthy features
of the more precise result.  First, the increase of the
light CP-even Higgs mass bound beyond $\mz$ can be significant.  This is
a consequence of the $m_t^4$ enhancement of the one-loop radiative
correction.
Second, the dependence of the light Higgs mass on the stop mixing
parameter $X_t$ implies that (for a given value of
$\msusy$) the upper bound of the light Higgs mass
initially increases with $X_t$ and reaches its {\it maximal} value
at $X_t=\sqrt{6}\msusy$.  This point is referred to as the {\it maximal
mixing} case (whereas $X_t=0$ corresponds to the {\it
minimal mixing} case). 

\begin{figure}[thb]
    \parbox{75mm}{\epsfxsize=\hsize\epsffile[82 207 550 568]
           {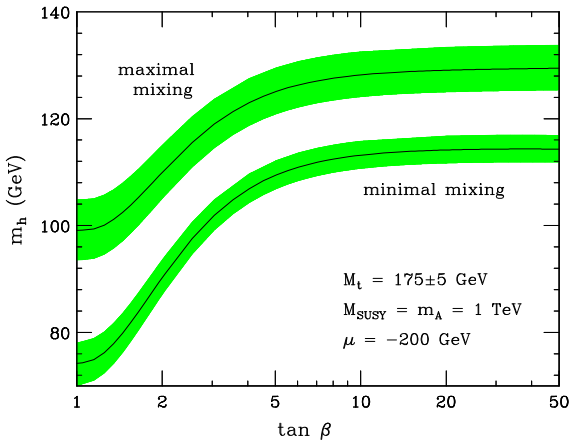}}
\caption[0]{The radiatively corrected
light CP-even Higgs mass is plotted
as a function of $\tanb$, 
for the maximal mixing [upper band] and minimal mixing cases.
The impact of the top quark
mass is exhibited by the shaded bands; the central value corresponds
to $M_t=175$~GeV, while the upper [lower] edge of the bands
correspond to increasing [decreasing] $M_t$ by 5~GeV.}
\label{mhtanb}
\end{figure}

Taking $\mha$ large, \fig{mhtanb} illustrates that the
maximal value of the lightest CP-even Higgs mass
bound is realized at large $\tanb$ in the case of maximal mixing.
Allowing for the
uncertainty in the measured value of $\mt$ and the
uncertainty inherent in the theoretical analysis,
one finds for $\MSUSY\lsim 2$~TeV that $\mhl\lsim
\mhmax$, where
\beqa \label{mhmaxvalue}
\mhmax&\simeq  122~{\rm GeV}, \quad
\mbox{minimal stop minimal,} \nonumber \\
\mhmax&\simeq  135~{\rm GeV}, \quad
\mbox{maximal stop mixing.}
\eeqa

The $\hl$ mass bound in the MSSM quoted above does not in general apply
to non-minimal supersymmetric extensions of the Standard Model.
If additional Higgs singlet and/or triplet fields are introduced,
then new Higgs self-coupling parameters appear, which are
not significantly constrained by present data.  For example, in the
simplest non-minimal supersymmetric extension of the
Standard Model (NMSSM),
the addition of a Higgs singlet adds a new Higgs
self-coupling parameter, $\lambda$ \cite{singlets}.
The mass of the lightest neutral Higgs boson can be
raised arbitrarily by increasing the value of $\lambda$ (analogous to
the behavior of the Higgs mass in the Standard Model!).
Under the assumption that all couplings stay perturbative
up to the Planck scale, one finds
in essentially all cases that $\mhl\lsim 200$~GeV, independent of the
details of the low-energy supersymmetric model~\cite{quiros}.

\subsection{MSSM Higgs searches at colliders}

There is presently no definitive experimental evidence for the Higgs boson.
Experimental limits on the charged and
neutral Higgs masses have been obtained at LEP.
For the charged Higgs boson, $\mhpm>78.7$~GeV \cite{LEPHIGGS}.
This is the most model-independent bound (it is valid for more general
non-supersymmetric two-Higgs doublet models) and assumes only that
the $H^+$ decays dominantly into $\tau^+\nu_\tau$ and/or $c \bar s$.
The LEP limits on the masses of $\hl$ and $\ha$ are obtained by searching
simultaneously for $e^+e^- \to Z \to Z\hl$ 
and $e^+e^- \to Z \to\hl\ha$.  At LEP1, the intermediate $Z$ is real
and the final state $Z$ is virtual, while at LEP2, the 
intermediate $Z$ is virtual and the final state $Z$ is real.
The $ZZ\hl$ and $Z\hl\ha$
couplings that govern these two decay rates are proportional to
$\sin(\beta-\alpha)$ and $\cos(\beta-\alpha)$, respectively.
Thus, one can use the LEP data to obtain simultaneous limits on
$\mhl$ and $\mha$, since the two tree-level masses determine $\alpha$
and $\beta$.   However, radiative corrections can be significant, 
so the final limits depend on the choice of MSSM
parameters that govern the radiative corrections (of which, the third
generation squark parameters are the most important). 
The present LEP 95\%~CL lower limits are $\mha>90.2$~GeV
and $\mhl>89.5$~GeV~\cite{LEPHIGGS}. 

The Higgs mass limits quoted above were
based on the assumption of $\msusy=1$~TeV and maximal 
stop mixing.\footnote{Although this tends to be a 
conservative assumption (that is,
other choices ensure that {\it more} of the $\mha$---$\tan\beta$ 
plane is covered), there are a number of other parameter regimes  
in which certain Higgs search strategies become more problematical.}
Under these assumptions, the
LEP MSSM Higgs search excludes the region of $0.53<\tanb<2.25$
At future colliders, the MSSM Higgs search will extend the excluded
region in the $\mha$---$\tanb$ plane.  Eventually, at least one
Higgs boson ($\hl$) must be discovered, or else low-energy
supersymmetry in its minimal form must be discarded.  It is possible
that additional Higgs bosons ($\hh$, $\ha$ and/or $\hpm$) will also be
discovered.  But, unlike $\hl$ whose mass is bounded from above
[\eq{mhmaxvalue}], the masses of $\hh$, $\ha$ and $\hpm$ are not so
restricted (although heavy Higgs boson masses must be below a few
TeV in order to preserve the naturalness of the theory).

In the region of MSSM Higgs parameter space where
$\mha\gg\mz$, one can show that
$\mhh\sim\mhpm\sim\mha$ [where the corresponding mass differences are
of ${\cal O}(\mz^2/\mha)$].  This parameter range is called {\it the
decoupling limit}~\cite{decoupling}, since it corresponds to the case where the
Higgs sector of the effective low-energy theory is equivalent to that
of the one-Higgs-doublet Standard Model.  In particular,
the properties of $\hl$ will be nearly indistinguishable from 
those of the Standard Model Higgs boson, while the heavier Higgs
bosons may be too heavy for discovery at the next generation of future
colliders. 

Let us consider briefly the prospects for discovering and making
precision measurements of the MSSM Higgs sector at
the upgraded Tevatron, LHC and the next $e^+e^-$ linear collider (LC).
The upgraded Tevatron begins running in 2001 at $\sqrt{s}=2$~TeV, 
with an initial goal of 
reaching 2~fb$^{-1}$ of integrated luminosity per year.  It has been suggested
that an ambitious program could achieve a
total integrated luminosity of 15~fb$^{-1}$ by the end of 2007.
LHC expect to begin taking data in 2006 at  $\sqrt{s}=14$~TeV, with an
initial goal of 10~fb$^{-1}$ 
per year.  Eventually, the
integrated luminosity is expected to reach 100~fb$^{-1}$ per year.
Finally, the next generation $e^+e^-$ LC is now currently under
development.  Initially, one expects the LC to operate at  
$\sqrt{s}=500$~GeV, with an integrated luminosity of 50~fb$^{-1}$ per year,
although there have been suggestions that the luminosity could be
improved by nearly an order of magnitude.

\begin{figure}[thb]
    \parbox{75mm}{\epsfxsize=\hsize\epsffile[0 0 567 567]
           {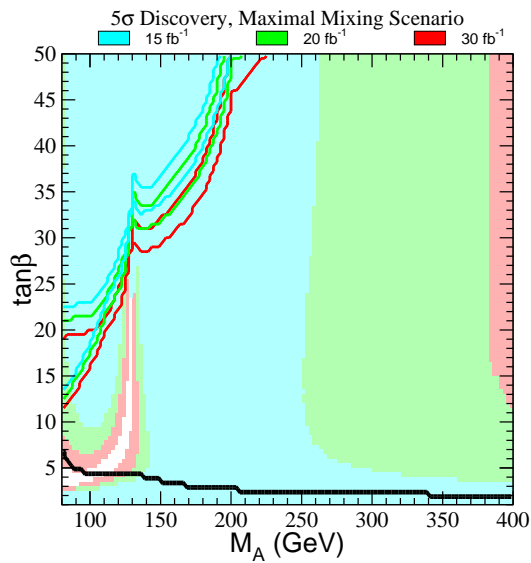}}
\caption[0]{\label{fullmhmax95} 
The anticipated 
Tevatron $5\sigma$ discovery region on the $\mha$---$\tan \beta$
plane, for the maximal mixing scenario
and two different search channels:
$q\bar q\to V\phi$ [$\phi=\hl$, $\hh$], $\phi\to b\bar b$
(shaded regions) and 
$gg$, $q\bar q\to b\bar b\phi$ [$\phi=\hl$, $\hh$, $\ha$],
$\phi\to b\bar b$ (region in the upper left-hand corner bounded by the
solid lines; the different lines correspond to CDF and D\O\
simulations).  Different integrated 
luminosities are explicitly shown by the color coding (or shades of gray).
The region below the solid black line near the bottom
of the plot is excluded by the absence of observed $e^+e^-\to Z\phi$
events at LEP2.  Taken from \Ref{tevreport}.
}
\end{figure}

The discovery reach of the Standard Model Higgs boson was analyzed by
the Tevatron Higgs Working Group in \Ref{tevreport}.   
The relevant production mechanism is $q_i\bar
q_j\to V\hsm$, where $V=W$ or $Z$. In all cases, it was assumed that
$\hsm$ would be observed via
$\hsm\to b\bar b$.  The most relevant final state signatures
corresponded to events in which the vector boson decayed leptonically
($W\to\ell\nu$, $Z\to\ell^+\ell^-$ and $Z\to\nu\bar\nu$, where
$\ell=e$ or $\mu$), resulting in $\ell\nu b\bar b$, $\nu\bar\nu b\bar
b$ and $\ell^+\ell^- b\bar b$ final states.  This analysis can  
be reinterpreted in terms of
the search for the CP-even Higgs boson of the MSSM.
In the MSSM at large $\tanb$, the enhancement of
the $\ha b\bar b$ coupling (and a similar enhancement of either the
$\hl b\bar b$ or $\hh b\bar b$ coupling), provides a new search
channel: $q\bar q$, $gg\to b\bar b\phi$ (where $\phi$ is a neutral
Higgs boson with enhanced couplings to $b\bar b$).  Combining both
sets of analyses, the Tevatron Higgs Working Group obtained the
anticipated 5$\sigma$ Higgs discovery contours for 
the maximal mixing scenario as a function of total integrated
luminosity per detector (combining both CDF and D\O\ data sets) shown in
\fig{fullmhmax95} \cite{tevreport}.

Fig.~\ref{fullmhmax95} shows that 
a total integrated luminosity of about 20~fb$^{-1}$ per experiment is 
necessary in order to assure a significant, although not exhaustive,
coverage of the MSSM parameter space.  If the anticipated 15~fb$^{-1}$
integrated luminosity is achieved, the discovery reach will
significantly extend beyond that of LEP. 
Nevertheless, the MSSM Higgs boson could still evade capture at the
Tevatron.  We would then turn to the LHC to try to obtain a definitive
Higgs boson discovery.

Over nearly\footnote{At the present writing,
not all possible MSSM parameter regimes have been studied with full
detector simulations.  Ref.~\cite{cmw} has pointed out that for certain
special choices of MSSM Higgs sector parameters, it may be very
difficult to achieve a $5\sigma$ discovery of {\it any} Higgs boson at
the LHC.  Moreover, it is important to note
that in other regions of the Higgs sector parameter space, the LHC
search strategies depend on the observation of small signals (above
significant Standard Model backgrounds) in more than one channel.
The present estimates of the statistical significance of the
Higgs signal rely on theoretical determinations of both signal and
background rates as well as simulations of detector 
performance.}
all the remaining region of the MSSM
Higgs sector parameter space, at least one of the MSSM Higgs bosons
will be detectable at the LHC, assuming that the machine runs at its design
luminosity of $100\pbi$ per year, and under the assumption that the
current detector design capabilities are achieved
\cite{lhchiggs}.
The LC would provide complete
coverage of the MSSM Higgs sector parameter space
(by extending the LEP-2 Higgs search) once the
center-of-mass energy of the machine is above 300~GeV
\cite{nlchiggs}.

However, complete coverage of the MSSM parameter space only means that
at least one Higgs boson of the MSSM
can be detected.  For example, in the
decoupling limit (where $\mha\gsim 2\mz$), $\hl$ will surely be
discovered at either the Tevatron or the LHC (and can be easily 
observed at the LC).  But, detection of the heavier
non-minimal Higgs states $\hh$, $\ha$, and/or $\hpm$
is not guaranteed.  At the LHC, there is a region \cite{lhchiggs}
in the $\mha$---$\tanb$ parameter space
characterized (very roughly) by $3\lsim\tanb\lsim 10$ and $\mha\gsim
300\gev$ such that only the $\hl$ will be detectable.\footnote{%
At large $\tanb$, the couplings of charged leptons to
$\ha$ (and $\hh$ in the decoupling limit)
are enhanced by a factor of $\tanb$.  Thus, gluon-gluon fusion to
$\ha$ or $\hh$ followed by Higgs decay to $\tau^+\tau^-$ and 
$\mu^+\mu^-$ should be observable at the LHC if $\tanb$ is large
enough (the value of the latter depends on $\mha$).}  The
LHC can make Higgs measurements (branching ratios and couplings)
with some precision~\cite{zeppen}, and
thus one can begin to check, if no other MSSM Higgs bosons are detected, 
whether the properties of $\hl$ 
deviate from those expected of the Standard Model Higgs boson ($\hsm$).

At the LC, $\hh\ha$ and $\hp\hm$
pair production are not kinematically allowed if $\mha\gsim \rts/2$.
Moreover, the production rate for $\hl\ha$
(although it may be kinematically allowed) is suppressed in the decoupling
limit.  Thus, the non-minimal Higgs states are
not directly detectable at the LC for $\mha\gsim \rts/2$.  
In the latter case, only the $\hl$ can be observed with properties
that are nearly identical to that of $\hsm$.  
More specifically, relative to the $\hsm$ couplings to fermion pairs, 
\beq
{g_{\hl f\bar f}\over g_{\hsm f\bar f}}=1+{\cal O}
\left({\mz^2\over\mha^2}\right)\,.
\eeq 
Thus, precision measurements of the $\hl$ branching ratios could
reveal small discrepancies that would provide an indication of the value
of $\mha$ even if the heavy MSSM Higgs boson cannot be kinematically
produced at the LC.  

Battaglia and Desch~\cite{battaglia}  
showed that for an LC
with $\sqrt{s}=350$~GeV and an ambitious
integrated luminosity of 500 fb$^{-1}$, one could measure
BR($\hl\to b\bar b$) to within an accuracy of about $\pm 2.5\%$.
To evaluate the significance of such a measurement, 
the percentage deviation of BR($\hl\to b\bar b$) in the MSSM relative
to that of $\hsm$ was computed in \Ref{chlm}.
The result of the computation is a set of
contours in the $\mha$---$\tanb$ plane, each one corresponding to a fixed 
percentage deviation of the BR.  The results are
sensitive to radiative corrections (which depend on the MSSM spectrum).
One would expect the deviation to vanish at large values of $\mha$
corresponding to the decoupling limit.  However, at large $\tanb$, the
approach to decoupling can be slowed due to $\tanb$-enhanced
radiative corrections.  
For example, for $\MSUSY=1$~TeV and maximal
stop mixing, \Ref{chlm} finds that 
the contour corresponding to a $3\%$ deviation in the BR
starts at $\mha\simeq  600$~GeV for $\tanb=3$ and slowly increases in $\mha$
until it reaches $\mha\simeq 1$~TeV for $\tanb=50$.  This means that a
precision measurement of BR($\hl\to b\bar b$) can provide 
evidence for the non-minimal Higgs sector, with some sensitvity to
large values of the heavy Higgs masses even if they lie above the
center-of-mass energy of the LC.

\section{THE LSP AND NLSP}

The phenomenology of low-energy supersymmetry depends crucially on
the properties of the lightest supersymmetric particle (LSP).
In R-parity conserving low-energy supersymmetry, all Standard Model
particles are R-even while their superpartners are R-odd.  Thus,
starting from an initial state involving
ordinary (R-even) particles, it follows that
supersymmetric particles must be
produced in pairs.  In general, these particles are highly unstable
and decay quickly into lighter states.  However, R-parity invariance
also implies that the LSP is absolutely
stable, and must eventually be produced
at the end of a decay chain initiated by the decay of a heavy unstable
supersymmetric particle.

In order to be
consistent with cosmological constraints, a stable LSP is almost
certainly electrically and color neutral \cite{Ellis84}.
Consequently, the LSP is weakly-interacting in ordinary
matter, \ie\ it behaves like a stable heavy neutrino
and will escape detectors without being directly observed.
Thus, the canonical signature for conventional R-parity-conserving
supersymmetric theories is missing (transverse)
energy, due to the escape of the LSP.

In mSUGRA models, the LSP is typically the lightest
neutralino, $\china$, and it tends to be dominated by its $\wt B$
component (an LSP with significant higgsino
components is possible only if $\tanb$ is large).
However, there are some regions of mSUGRA parameter space
where other possibilities for the LSP are realized.
For example, there are regions of mSUGRA parameter space where
the LSP is a chargino.  These regions must be excluded since we reject
the possibility of charged relic particles surviving the early
universe \cite{champs}.
In addition, one may impose cosmological constraints (such that the
relic LSP's do not ``overclose'' the universe by
contributing a mass density that is larger than
the critical density of the universe \cite{griest}) to rule out
additional regions of mSUGRA parameter space.  The condition
that the relic density of LSP's constitutes a significant part of the
dark matter imposes even further restrictions on the mSUGRA parameter
space \cite{susydm}.

In more general SUGRA
models, the nature of the LSP need not be so constrained.  One can
envision a $\china$-LSP which has an arbitrary mixture of gaugino and
higgsino components by relaxing gaugino mass unification.
A nearly pure higgsino LSP is
possible in the region where the (low-energy) gaugino Majorana mass
parameters satisfy
$M_1\simeq M_2\gsim\mu$ \cite{hlsp}.  The sneutrino
(in particular, the $\snu_\tau$) can be a viable LSP,
although it is unlikely to be a major component of the dark matter
\cite{srednicki} unless the supersymmetric model incorporates some
lepton number violation \cite{snudm}.
In AMSB models, \eq{eqanom} yields
$M_1\simeq 2.8M_2$ and $M_3\simeq -8.3M_2$, which implies that the
lightest chargino pair and neutralino make up a nearly
mass-degenerate triplet of winos; the corresponding LSP is
approximately a pure $\wt W^3$.

In most GMSB models,
the mass of the gravitino lies in the eV---keV regime.
Thus, in this scenario
the gravitino will be the LSP and the next-to-lightest
supersymmetric particle (NLSP) also plays a crucial role in the phenomenology
of supersymmetric particle production and decay.  Note that unlike the
LSP, the NLSP can be charged.   In GMSB models,
the most likely candidates for
the NLSP
are $\china$ and $\wt\tau_R^\pm$.
The NLSP will decay into its Standard Model superpartner plus a
gravitino
[either  $\china\to N\gtino$ ($N=\gamma$, $Z$, or $\hl$) or
$\wt\tau_R^\pm\to\tau^\pm\gtino $], with a lifetime that is
quite sensitive to the model parameters.
In a small range of parameter space, it is possible that several of the
next-to-lightest supersymmetric particles are sufficiently degenerate
in mass such that each one behaves as the NLSP.\footnote{For example, if
$\wt\tau_R^\pm$ and $\china$ are nearly degenerate in mass, then neither
$\wt\tau_R^\pm\to\tau^\pm\china$ nor $\china\to\wt\tau_R^\pm\tau^\mp$
are kinematically allowed decays.  In this case, $\wt\tau_R^\pm$ and
$\china$ are co-NLSP's, and each decays dominantly into its Standard
Model superpartner plus a gravitino.}
In this case, these particles are called co-NLSP's
\cite{conlsp}. Different
choices for the identity of the NLSP and its decay rate lead to a
variety of distinctive supersymmetric phenomenologies
\cite{thomas,eegamgamkane}.

Since a light gravitino is stable in R-parity conserving
GMSB models, it is also a candidate for dark
matter \cite{primack}.  Although very light gravitinos (eV masses)
will not contribute significantly to the total mass density of the
universe, the requirement that their relic density
not overclose the universe implies that $\mgtino\lsim$ few keV in the
usual early universe scenarios.  Alternative scenarios do exist in which
the gravitino of GMSB models can be somewhat heavier.  In SUGRA-based
models, the gravitino is typically not the LSP and thus is unstable,
although its lifetime is quite long and
is relevant only in considerations of early universe cosmology.
The pertinent issues are nicely summarized in \Ref{gmsbreview}.

More generally, in any R-parity-conserving supersymmetric model
it is important to check that the relic density of LSP's
does not overclose the universe. This can
lead to useful constraints on the parameters of the
model.
Even if the LSP is a viable dark matter candidate,
one should not necessarily constrain the
parameters of the MSSM by requiring the LSP to be a major component of
the dark matter. It may turn out that the main component of the
dark matter has another source.  Some examples are:
the QCD axion, its supersymmetric partner (the axino \cite{axino})
or the lightest stable particle in the GMSB messenger sector
\cite{taohan}.

\section{CLASSES OF SUPERSYMMETRIC SIGNALS AT FUTURE COLLIDERS}

The lack of knowledge of the origin and structure of the
SUSY-breaking parameters implies that the predictions for low-energy
supersymmetry and the consequent phenomenology
depend on a plethora of unknown parameters.
Many details of supersymmetric phenomenology are
strongly dependent on the underlying assumptions of the model.
Nevertheless, one can broadly classify supersymmetric
signals at future colliders by considering the various theoretical approaches
described in Section~3.

\subsection{Missing energy signatures}

In R-parity conserving low-energy supersymmetry, supersymmetric particles are
produced in pairs.    The subsequent decay of a heavy
supersymmetric particle generally proceeds via a multistep decay chain
\cite{leveille,chain,enhancedbs}, ending in the production of at least
one supersymmetric particle that (in conventional models)
is weakly interacting and escapes the
collider detector.  Thus, supersymmetric particle production yields events that
contain at least two escaping non-interacting particles, leading to a missing
energy signature.   At hadron colliders, it is only possible to detect missing
transverse energy ($\etmiss$),
since the center-of-mass energy of the hard collision
is not known on an event-by-event basis.

In conventional SUGRA-based models, the weakly-interacting LSP's that
escape the collider detector (which yields large missing transverse
energy) are accompanied by energetic jets and/or
leptons.  This is the ``smoking-gun'' signature of low-energy supersymmetry.
In contrast, in AMSB models,
the $\china$ is the LSP but the lightest
neutralino and chargino are nearly degenerate in mass.  If the mass difference
is $\lsim 100$~MeV, then $\chipa$ is long-lived and decays outside the detector
\cite{guniondrees,Feng99}.  In this case, some supersymmetric
events would yield {\it no} missing energy and two semi-stable charged
particles that pass through the detector.   

In conventional GMSB models with a gravitino-LSP, all supersymmetric
events contain at least two NLSP's, and the resulting signature depends on the
NLSP properties.   Four physically distinct possible scenarios emerge:
\begin{itemize}
\item
The NLSP is electrically and color neutral and long-lived, and
decays outside of the detector
to its associated Standard Model partner and the gravitino.
\item
The NLSP is the sneutrino and decays invisibly
into $\nu \gtino$ either inside or outside the detector.
\end{itemize}
In either of these two cases, the resulting missing-energy signal is
then similar to that of the SUGRA-based models where $\china$ or
$\snu$ is the LSP.
\begin{itemize}
\item
The NLSP is the $\china$ and decays inside the detector to $N \gtino$, where
$N=\gamma$, $Z$ or a neutral Higgs boson.
\end{itemize}
In this case, the gravitino-LSP behaves like the neutralino or sneutrino LSP of
the SUGRA-based models.  However, 
the missing energy events of the GMSB-based model are
characterized by the associated production of (at least) two $N$'s, one for
each NLSP.
Note that if $\china$ is lighter than the $Z$ and
$\hl$ then BR$(\china\to\gamma \gtino)=100\%$, and all supersymmetric
production will result in missing energy events with at least two associated
photons.
\begin{itemize}
\item
The NLSP is a charged slepton (typically $\staur$ in GMSB models
if $\mstaur<\mcnone$),
which decays to the corresponding lepton partner and gravitino.
\end{itemize}
If the decay is prompt, then one finds missing energy events with associated
leptons.  If the decay is not prompt, one observes a long-lived
heavy semi-stable charged particle with {\it no} associated missing energy
(prior to the decay of the NLSP).

There are also GMSB scenarios in which there are several nearly degenerate
so-called co-NLSP's, any one of which can
be produced at the penultimate step of the supersymmetric decay
chain. The resulting supersymmetric signals would consist of events with
two (or more) co-NLSP's, each one of which would decay according to
one of the four scenarios delineated above.  For additional details on
the phenomenology of the co-NLSP's, see  \Ref{conlsp}.

In R-parity violating SUGRA-based models the LSP is unstable.  If the
RPV-couplings are sufficiently weak, then the LSP will decay outside the
detector, and the standard missing energy signal applies.  If the LSP decays
inside the detector, the phenomenology of RPV models depends on the
identity of the LSP and the branching ratio of possible final state
decay products. If the latter includes a neutrino, then the corresponding
RPV supersymmetric
events would result in missing energy (through neutrino emission) in
association with hadron jets and/or leptons.  However, other decay chains are
possible depending on the relative strengths of $\lambda_L$, $\lambda^\prime_L$
and $\lambda_B$ [see \eq{nrsuppot}].  Other possibilities include decays into
charged leptons in association with jets (with no neutrinos), and decays into
purely hadronic final states.  Clearly, these latter events would
contain little missing energy. If R-parity violation is
present in GMSB models, the RPV decays of the NLSP can easily dominate
over the NLSP decay to the gravitino. In this case, the phenomenology
of the NLSP resembles that of the LSP of SUGRA-based RPV 
models~\cite{dreiner}.

\subsection{Lepton ($e$, $\mu$ and $\tau$) signatures}

Once supersymmetric particles are produced at colliders, they do not
necessarily decay to the LSP (or NLSP) in one step.  The resulting decay chains
can be complex, with a number of steps from the initial decay to the final
state \cite{chain}.
Along the way, decays can produce real or virtual $W$'s, $Z$`s,
charginos, neutralinos and
sleptons, which then can produce leptons in their subsequent decays.  Thus,
many models yield large numbers of
supersymmetric events characterized by one or more leptons in
association with missing energy, with or without hadronic jets.

One signature of particular note is events containing like-sign
di-leptons \cite{likesign}.
The origin of such events is associated with the Majorana nature of the
gaugino.  For example, $\gl\gl$ production, followed by gluino decay via
\beq
\gl\to q\anti q\cpmone\to
q\anti q\ell^\pm\nu\cnone
\label{gldecay}
\eeq
can result in like-sign leptons since the $\gl$ decay leads with equal
probability to either $\ell^+$ or $\ell^-$.
If the masses and mass differences are both substantial
(which is typical in mSUGRA models, for example), like-sign di-lepton
events will be characterized by fairly energetic jets and isolated
leptons and by large $\etmiss$ from the LSP's.
Other like-sign di-lepton signatures can arise in a similar way from the decay
chains initiated by the heavier neutralinos.

Distinctive tri-lepton signals~\cite{bcpttri} can result from
$\cpmone\cntwo\to(\ell^{\pm}\nu\cnone)(\ell^+\ell^-\cnone)$.
Such events have little hadronic activity (apart
from initial state radiation of jets off
the annihilating quarks at hadron
colliders).  These events can have a variety of interesting characteristics
depending on the fate of the final state neutralinos.

If the soft-SUSY-breaking slepton masses are
flavor universal at the high energy scale $M_X$ (as in mSUGRA models)
and $\tanb\gg 1$, then
the $\staur$ will be significantly lighter than the other slepton states.
As a result, supersymmetric decay chains involving (s)leptons will
favor $\staur$ production, leading to a predominance of events with
multiple $\tau$-leptons in the final state.

In GMSB models with a charged slepton NLSP, the decay
$\slep\to\ell\,\gtino$ (if prompt)
yields at least two leptons for every supersymmetric event
in association with missing energy.
In particular, in models with a $\staur$ NLSP, supersymmetric events
will characteristically contain at least two $\tau$'s.

In RPV models, decays of the LSP
(in SUGRA models) or NLSP (in GMSB models)
mediated by RPV-interactions proportional to $\lambda_L$ and $\lambda^\prime_L$
will also yield supersymmetric events containing charged leptons. However,
if the only significant RPV-interaction is the one proportional to
$\lambda^\prime_L$, then such events would
contain little missing energy (in contrast to the GMSB signature
described above).

\subsection{$b$-quark signatures}

The phenomenology of gluinos and squarks depends critically on their relative
masses.  If the gluino is heavier, it will decay dominantly into
$q\sq$,\footnote{In the following, the notation $q\sq$ means
either $q\anti{\sq}$ or $\anti{q}\sq$.}
while the squark can decay into quark plus chargino or
neutralino.  If the squark is heavier, it will decay dominantly into a
quark plus gluino, while the gluino will decay into the three-body modes
$q\bar q\wt\chi$ (where $\wt\chi$ can be either a neutralino or
chargino, depending
on the charge of the final state quarks).  A number of special cases can arise
when the possible mass splitting among squarks of different flavors is taken
into account.  For example, models of supersymmetric mass spectra have been
considered where the third generation squarks are lighter than the squarks of
the first two generations.  If the gluino is lighter than the latter but
heavier than the former, then the only open gluino two-body decay mode
could be $b\sbot$.  In such a case, all $\gl\gl$ events will result in at least
four $b$-quarks in the final state (in associated with the usual missing energy
signal, if appropriate).  More generally, due to the flavor
independence of the strong interactions, one expects three-body
gluino decays into
$b$-quarks in at least 20\% of all gluino decays.\footnote{Here we assume the
approximate degeneracy of the first two
generations of squarks, as suggested from the
absence of FCNC decays.  In many models, the $b$-squarks tend to
be of similar mass or lighter than the squarks of the first two generations.}
Additional $b$-quarks can
arise from both top-quark and top-squark decays,
and from neutral Higgs bosons produced somewhere in the
chain decays \cite{bquarkbaer}.
Finally, at large $\tanb$, the enhanced Yukawa coupling to $b$-quarks
can increase the rate of $b$-quark production in
neutralino and chargino decays occurring at some step in the gluino
chain decay.
These observations suggest that many supersymmetric events at hadron colliders
will be characterized by $b$-jets in association with missing
energy~\cite{enhancedbs,bquarkian}.

\subsection{Signatures involving photons}

In mSUGRA models, most supersymmetric events do not contain isolated
energetic photons.  However, some areas of low-energy supersymmetric parameter
space do exist in which final state photons can arise in the decay chains of
supersymmetric particles.  If one relaxes the condition of gaugino
mass unification [\eq{gunif}], interesting alternative supersymmetric
phenomenologies can arise.  For example, if $M_1\simeq M_2$, then the branching
ratio for $\chinb\to\china\gamma$ can be significant \cite{wyler}.
In the model of
\Ref{eegamgamkane}, the $\china$-LSP is dominantly higgsino, while $\chinb$ is
dominantly gaugino.  Thus, many supersymmetric decay chains end in the
production of $\chinb$, which then decays to $\china\gamma$.  In this picture,
the pair production of supersymmetric particles often
yields two photons plus associated missing energy.  

In GMSB models with a $\china$-NLSP, all supersymmetric decay chains would end
up with the production of $\china$.  Assuming that $\china$ decays inside the
collider detector, one possible decay mode is $\china\to\gamma \gtino$.
In many models, the branching ratio for this
radiative decay is significant (and could be as
high as 100\% if other possible two-body decay modes are not kinematically
allowed).  In the latter case, supersymmetric pair production would also yield
events with two photons in associated with large missing energy.  The
characteristics of these events differ in detail from those of
the corresponding events expected in the model of \Ref{eegamgamkane}.

\subsection{Kinks and long-lived heavy particles}

In most SUGRA-based models, all supersymmetric particles in the decay chain
decay promptly until the LSP is reached.  The LSP is exactly stable and
escapes the collider detector.  However, exceptions are possible.
In particular, if there is a supersymmetric
particle that is just barely heavier
than the LSP, then its (three-body) decay rate
to the LSP will be significantly suppressed and it could be long lived.  For
example, in AMSB models where
$\mcpmone\simeq\mcnone$, the $\cpmone$ can be sufficiently long
lived to yield a detectable vertex, or perhaps even exit the 
detector~\cite{guniondrees,Feng99}.

In GMSB models, the NLSP may be long-lived, depending on its mass and
the scale of SUSY-breaking, $\sqrt{F}$.
The couplings of the NLSP to the
helicity $\pm\half$ components of the gravitino
are fixed by \eq{gtrel}.
For $\sqrt F\sim 100$---$10^4\tev$, this coupling is very weak,
implying that all the supersymmetric particles other than the NLSP undergo
chain decays down to the NLSP (the branching ratio for the direct
decay to the gravitino is negligible). The NLSP is unstable and
eventually decays to the gravitino.
For example, in the case of the $\china$-NLSP (which is dominated by
its $\wt B$ component), one can use \eq{gtrel}\ to obtain
$\Gamma(\cnone\to \gam\gtino)=
m^5_{\tilde\chi_1^0}\cos^2\theta_W/16\pi F^2$.  It then follows that
\begin{equation}
(c\tau)_{\cnone}\simeq 130
\left({100\gev\over\mcnone}\right)^5
\left({\sqrt F\over 100\tev}\right)^4\mu {\rm m}\,.
\label{ctauform}
\end{equation}

For simplicity, assume that $\china\to\gamma\gtino$ is the dominant NLSP
decay mode.  If $\sqrt F\sim
10^4\tev$, then the decay length for the NLSP is
$c\tau\sim 10$~km for $\mcnone=100\gev$; while
$\sqrt F\sim 100\tev$ implies a short but vertexable decay length.
A similar result is obtained in the case of a charged NLSP.  Thus, if
$\sqrt{F}$ is sufficiently large, the charged NLSP will be semi-stable and
may decay outside of the collider detector.

Finally, if R-parity violation is present,
the decay rate of the LSP in SUGRA-based models
(or the NLSP in R-parity-violating GMSB models)
could be in the relevant range
to yield visible secondary vertices.

\subsection{Exotic supersymmetric signatures}

If R-parity is not conserved, supersymmetric phenomenology exhibits
many features that are quite distinct from those of the MSSM~\cite{dreiner}.
Both $\Delta L\!=\! 1$ and
$\Delta L\!=\! 2$ phenomena are allowed (if L is
violated), leading to neutrino masses
and mixing \cite{numass}, 
neutrinoless double beta decay \cite{0nubb},
and sneutrino-antisneutrino mixing \cite{grosshab}.
Further, since the distinction between the Higgs and matter
multiplets is lost, R-parity violation permits the mixing of sleptons
and Higgs bosons, the mixing of neutrinos and neutralinos, and the
mixing of charged leptons and charginos, leading to more complicated
mass matrices and mass eigenstates than in the MSSM.

Some of the consequences for collider signatures have already
been mentioned.  Most important, the LSP in RPV SUGRA models
is no
longer stable, which implies that not all supersymmetric decay chains
must yield missing-energy events at colliders.
In particular, if $\china$ is the LSP, then its
RPV decays contain visible particles:
\beq
\cnone\to \underbrace{(jjj)}_{\lam_B\neq 0},
~~\underbrace{(\ell\ell^{(\prime)}\nu)}_{\lam_L\neq 0},
~~\underbrace{(\ell jj,\nu jj)}_{\lam_L^\prime\neq0}\,,
\label{cnonedecays}
\eeq
where $j$ stands for a hadronic jet (in this case arising from single
quarks or anti-quarks), and the relevant RPV-coupling
[in the notation of \eq{nrsuppot}] is indicated below
the corresponding channel.  Thus, even $\epem\to \cnone\cnone$ pair
production and sneutrino decay via
$\snu\to \nu\cnone$ become visible (since the $\cnone$ now decays
into visible channels).  Likewise, a sneutrino LSP would also yield
visible decay modes.  For example, $\snu$---$\hl$ mixing could lead to a
substantial $\snu\to b\bar b$ decay branching ratio.  The relative
strength of $\snu\to q\bar q$, $\ell^+\ell^-$ also depends on the
strength of the L-violating RPV couplings.  In addition, $\china\to
\bar\nu\snu$ (or its charge-conjugated state) would be visible once
the sneutrino decayed.

A number of other phenomenological consequences are noteworthy.
For $\lambda_L\neq 0$ sneutrino resonance production in
$\epem$ \cite{schannel} collisions becomes
possible.  For $\lambda_L^\prime\neq 0$, squarks can be regarded as
leptoquarks~\cite{butterworth} 
since the following processes are allowed:
$e^+\overline u_m\to \overline{\widetilde d}_n\to e^+\overline
u_m$, $\overline\nu\overline d_m$ and
$e^+ d_m \to \widetilde u_n\to e^+d_m$ (where
$m$ and $n$ are generation labels).
The same term responsible for the processes displayed above could also generate
purely hadronic decays for sleptons and sneutrinos:
{\it e.g.}, $\widetilde\ell^-_p\to
\overline u_m d_n$ and $\widetilde\nu_p\to \overline q_m q_n$ ($q=u$ or
$d$).  If such decays were dominant, then the pair production of
sleptons in $e^+e^-$ events would lead to hadronic four-jet
events with jet pairs of equal mass \cite{fourjet}, a
signature quite different from the missing energy signals expected in
the MSSM. Alternatively, $\lambda_L\neq 0$ could result in
substantial branching fractions for
$\slep\to \ell\nu$ and $\snu\to \ell^+\ell^-$ decays.
Sneutrino pair production would then yield
events containing four charged leptons with two lepton pairs of equal mass.

\section{CONCLUSIONS}

Low-energy supersymmetry remains the most elegant solution to the
naturalness and hierarchy problems, while
providing a possible link to Planck scale physics and the unification of
particle physics and gravity.  Nevertheless, the
origin of the soft supersymmetry-breaking
terms and the details of their structure remain a mystery.
There are many theoretical ideas, but we still cannot be certain which
region of the MSSM-124 parameter space (or some non-minimal extension
thereof) is the one favored by nature.
The key theoretical breakthroughs will surely
require experimental guidance and input.

Thus, we must rely on experiments at future colliders
to uncover evidence for low-energy supersymmetry.
Canonical supersymmetric signatures at
future colliders are well analyzed and understood.  Much of the recent
efforts have been directed at trying to
develop strategies for
precision measurements to
establish the underlying supersymmetric structure
of the interactions and to distinguish among models.  However, we
are far from understanding all possible facets of MSSM-124 parameter
space (even restricted to those regions that are phenomenologically
viable).  For example, the phenomenological implications of the
potentially new CP-violating phases that can arise in the MSSM and its
consequences for collider physics have only recently begun to attract
attention.  Moreover, the variety 
of possible non-minimal models of low-energy supersymmetry
presents additional challenges to experimenters who plan on searching
for supersymmetry at future colliders.

If supersymmetry is discovered, it will provide a plethora of
experimental signals and theoretical analyses.
The variety of phenomenological manifestations
and parameters of supersymmetry suggest that many years of experimental
and theoretical work will be required before it will be possible to determine
the precise nature of supersymmetry-breaking and the
implications for a more fundamental theory of particle interactions.

\section*{ACKNOWLEDGMENTS}

I would like to gratefully acknowledge the collaboration of Jack
Gunion in preparing sections 5 and 6 of this paper.

\end{document}